\documentclass{ws-procs975x65}

\usepackage[T1]{fontenc}
\usepackage[utf8]{inputenc}
\usepackage{lmodern}
\usepackage[english]{babel}

\usepackage{graphicx}

\newcommand{\pd}[2]{\frac{\partial #1}{\partial #2}}

\begin{document}
\title{COUPLING DIMERS TO CDT - CONCEPTUAL ISSUES}

\author{LISA GLASER}

\address{The Niels Bohr Institute, Copenhagen University,\\
Blegdamsvej 17, DK-2100 Copenhagen \O , Denmark.\\
E-mail: glaser@nbi.dk }



\begin{abstract}
Causal dynamical triangulations allows for a non perturbative approach to quantum gravity.
In this article a solution for dimers coupled to CDT is presented and some of the conceptual problems that arise are reflected upon.
\end{abstract}

\bodymatter
\section{Introduction}
Causal Dynamical Triangulations (CDT) is a proposed theory of quantum gravity. As in dynamical triangulations the path integral for gravity is regularized through simplices. In CDT a preferred time slicing is introduced to allow a well-defined Wick rotation. This preferred time slicing leads to a better behaved continuum theory \cite{ambjorn98}.

CDT in two dimensions can also be solved analytically using matrix models \cite{ambjorn_matrix_2008}.
It is well-know that random lattices can be coupled to matter, like dimers or the Ising model, and that this leads to quantum gravity coupled to conformal field theories \cite{kazakov_appearance_1989,matthias_yang-lee_1990}. It is then an interesting prospect to try and couple matter to the random lattices of CDT.

In this proceedings contribution we will very shortly introduce how one can solve dimers coupled to CDT and then elaborate on some of the problems encountered in doing so\cite{ambjorn_new_2012,Atkin:2012yt}.

\section{CDT as a rooted tree}
\begin{figure}
\centering
\begin{minipage}{.45\textwidth}
  \centering
\includegraphics[width=0.55\linewidth]{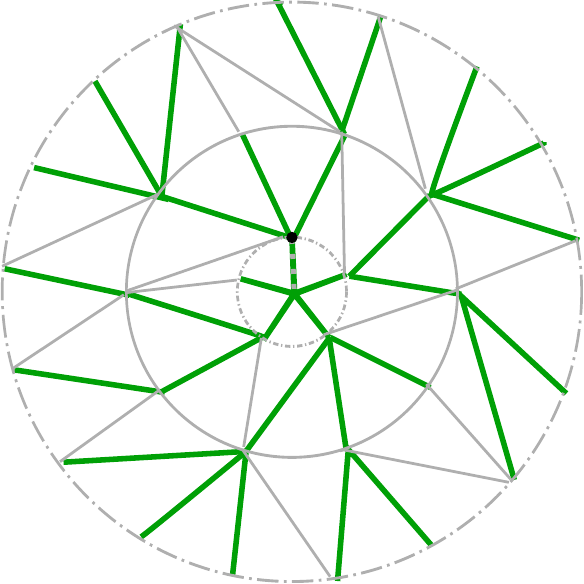}
\end{minipage}\hfill
\begin{minipage}{.45\textwidth}
  \centering
\includegraphics[width=0.9\linewidth]{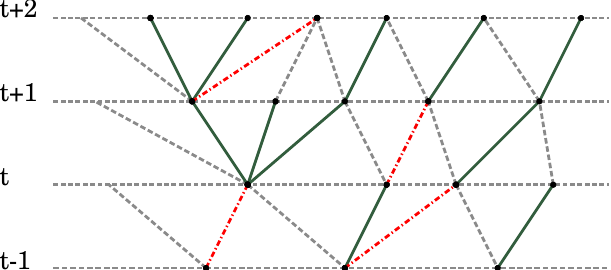}
\end{minipage}
\caption{\label{fig:bijection}On the left is a embedding of a simple CDT. The green marked lines on it are those that are also part of the graph. These suffice to characterize the entire CDT. If we only had the green lines we could get back to the full CDT by just reintroducing the space-like links and the leftmost link at every vertex.
The figure on the right are three time slices of a CDT. The tree graph is marked green. The red markings indicate dimers, which can be placed on the tree so as to be non touching.
}
\end{figure}

Durhuus et al\cite{durhuus09} proved that there is a bijective mapping between rooted tree graphs (cf. fig \ref{fig:bijection}) and CDT.
This bijection makes it possible to determine the critical exponents of CDT using recursive equations \cite{QuGeom}. 

Dimers are a simple matter model. A dimer can be described as a mark that is placed on a link in the rooted tree. The expression hard dimer means that dimers may not be placed on adjacent links. Then the simple rule of placing any number of hard dimers on the tree will lead to a partition function which allows for new multicritical behavior\cite{ambjorn_new_2012,glaser_Prague_2012}.

There have been attempts to couple dimers to CDT before; Francesco et al. \cite{kristjansen99} introduced a model where dimers were placed on the space like links only. Their dimer model allows for treatment with a transfer matrix approach, but is found to be in the same universality class as ordinary CDT.

This can be understood considering that such dimers do not create interaction between the time like slices of the CDT. Thus any modification of the triangulation is strictly time-local at every instant and can not change the global structure of the triangulation.
Dimers only placed on the tree graph on the other hand
always create interaction between different time slices.

One restriction that placing dimers only on the tree leads to is that no dimers can be placed on the leftmost link of a vertex. Removing this restriction would break the bijective mapping, but it seems unlikely for it to change the universality class of the solution.
\section{Dimers and multicriticality}
After dimers have been introduced we can write the partition function as
\begin{equation}
Z(\mu ,\xi ) = \sum_{BP}  e^{-\mu } \sum_{{\rm HD(BP)}} \xi^{|{\rm HD(BP)}|} \quad.
\end{equation}
where $\xi$ is the weight associated with a dimer, $BP$ denotes the ensemble of tree graphs, and ${\rm HD(BP)}$ the different dimer configurations on each tree graph. 
This partition function can be calculated using recursive relations arising for the rooted tree graphs.
One finds the coupled set of equations
\begin{align}
Z&= e^{-\mu} \left( \frac{1}{1-Z} +  W \frac{1}{(1-Z)^2} \right) &	W&= e^{-\mu}\xi \left( \frac{1}{1-Z} \right) \quad ,
\end{align}
where $Z$ is the partition function for a tree with a normal link at the root and $W$ is the partition function for a tree rooted in a dimer. The condition for multicriticality is that
\begin{equation}
\pd{\mu(Z,\zeta_c)}{Z}\bigg|_{Z_c} = \pd{^2 \mu(Z,\zeta_c)}{Z^2}\bigg|_{Z_c}=0 \;
\end{equation}
this is satisfied for
		$Z_c=   \frac{5}{8} , \xi_c= 	-\frac{1}{12} , e^{\mu_c}=  \frac{32}{9}$.
The critical exponents can then be calculated,
the string susceptibility $\gamma =\frac{1}{3}$, the Hausdorff-dimension $d_H=\frac{3}{2}$ and the edge singularity $\sigma= \frac{1}{2}$.
The multicritical point lies at a value of $\xi <0$. This is slightly troubling since we ordinarily would interpret the weight as being related to the probability of the state.

  
A possible interpretation is to allow negative probabilities just as conditional probabilities in a Bayesian framework\cite{haug}. In our model that would mean that the probability to find a tree decorated with an odd number of dimers would be negative, but the sum of the probabilities for odd and even numbers of dimers, so the probability to find a tree independent of the dimers on it would still be positive. This interpretation seemed promising, so we tried to formalize it.
Numerical for trees of up to 6 links it is satisfied, however  so far the search for an analytic expression remained without success.

The negative probabilities also make it difficult to decide what type of configuration is typical at the critical point. It is not possible to run Monte Carlo simulations to compare to the analytic solution for our dimer model. 
The Bayesian interpretation might allow for simulations, however as of yet this has not been pursued.
\section{Summary}
Coupling dimers to CDT is related to a theory of non unitary matter coupled to gravity. It can be solved and the critical exponents show clearly that it lies in a different universality class than pure CDT.

But how to think about this matter in more physical terms is puzzling. The concept of dimers with a negative weight defies physical intuition.
It might be promising to think about these negative weights as Bayesian probabilities and only consider the sum over all dimer configurations on the same tree as a physical quantity. This might then be used to simulate systems with dimers to find a better intuition about this type of system.

\section*{Acknowledgments}
I would like to thank the Danish Research Council for financial support via the grant ``Quantum gravity and the role of Black holes''. 

\bibliographystyle{ws-procs975x65}
\bibliography{bibliography}
\end{document}